\documentstyle[12pt]{article}
\begin{document}
%\addtolength{\baselineskip}{.05mm}
\input epsf
\newcommand{\ttau}{r}
\newcommand{\vev}[1]{\langle #1 \rangle}
\def\mapright#1{\!\!\!\smash{
\mathop{\longrightarrow}\limits^{#1}}\!\!\!}
\newcommand{\bigoint}{\displaystyle \oint}
\newlength{\extraspace}
\setlength{\extraspace}{3mm}
\newlength{\extraspaces}
\setlength{\extraspaces}{4mm}
\newcounter{dummy}
\newcommand{\be}{\begin{equation}
\addtolength{\abovedisplayskip}{\extraspaces}
\addtolength{\belowdisplayskip}{\extraspaces}
\addtolength{\abovedisplayshortskip}{\extraspace}
\addtolength{\belowdisplayshortskip}{\extraspace}}
\newcommand{\ee}{\end{equation}}
\newcommand{\newsection}[1]{
\vspace{18mm}
\pagebreak[3]
\addtocounter{section}{1}
\setcounter{subsection}{0}
\setcounter{footnote}{0}
\noindent
{\large\bf \thesection. #1}
\nopagebreak
\bigskip
\nopagebreak}

\newcommand{\itsubsection}[1]{
\vspace{1cm}
\pagebreak[3]
\addtocounter{subsection}{1}
\addcontentsline{toc}{subsection}{\protect
\numberline{\arabic{section}.\arabic{subsection}}{#1}}
\noindent{\bf \thesection.\thesubsection. #1}
\nopagebreak
\vspace{2mm}
\nopagebreak}
\newcommand{\ba}{\begin{eqnarray}
\addtolength{\abovedisplayskip}{\extraspaces}
\addtolength{\belowdisplayskip}{\extraspaces}
\addtolength{\abovedisplayshortskip}{\extraspace}
\addtolength{\belowdisplayshortskip}{\extraspace}}
\newcommand{\FF}{{\cal F}}
\newcommand{\one}{{\bf 1}}
\newcommand{\zbar}{\overline{z}}
\newcommand{\ea}{\end{eqnarray}}
\newcommand{\is}{& \!\! = \!\! &}
\newcommand{\hf}{{1\over 2}}
\newcommand{\del}{\partial}
%
% 2 by 2 matrices
%
\newcommand{\twomatrix}[4]{{\left(\begin{array}{cc}#1 & #2\\
#3 & #4 \end{array}\right)}}
\newcommand{\twomatrixd}[4]{{\left(\begin{array}{cc}
\displaystyle #1 & \displaystyle #2\\[2mm]
\displaystyle  #3  & \displaystyle #4 \end{array}\right)}}
\newcommand{\low}{{{\rm {}_{IR}}}}
\newcommand{\hi}{{{\rm {}_{UV}}}}
\newcommand{\hilo}{{{}_{{}^{\rm {}_{UV/IR}}}}}
\newcommand{\ie}{{\it i.e.\ }}
\newcommand{\gbar}{{\overline{g}}}
\newcommand{\half}{{\textstyle{1\over 2}}}
\newcommand{\tfrac}{\frac}
\newcommand{\XX}{\dot{g}}
\newcommand{\XXX}{{\mbox{\small \sc X}}}
\newcommand{\xx}{p}
\newcommand{\kappaf}{\kappa_{{}_{\! 5}}}
\newcommand{\CCC}{{\mbox{\large $\gamma$}}}
\renewcommand{\thesubsection}{\arabic{subsection}}
\renewcommand{\footnotesize}{\small}
\newcommand{\figuur}[3]{\begin{figure}[t]\begin{center}
\leavevmode\hbox{\epsfxsize=#2 \epsffile{#1.eps}}\\[3mm] \bigskip
\parbox{15.5cm}{\small \bf Fig.1.\ \it  #3} \end{center}
\end{figure}\hspace{-1.5mm} }
\newcommand{\figuren}[3]{
 \begin{figure}[t]\begin{center}
\leavevmode\hbox{\epsfxsize=#2 \epsffile{#1.eps}}\\[3mm] \bigskip
\parbox{15.5cm}{\small \bf Fig.2.\ \it  #3} \end{center}
\end{figure}\hspace{-1.5mm}}

\begin{titlepage}

{\hbox to\hsize{ \hfill August 2000}} \vspace{2cm}

 \begin{center}

 {\LARGE On the Holographic Principle in} \\[.8cm]
 {\LARGE{a Radiation Dominated Universe}}

 \vspace{2.6cm}

 {\large \sc{Erik Verlinde}}\\[.6cm]

 {\it{Joseph Henry Laboratories}} {\it{\ Princeton University}}

 {\it{Princeton, New Jersey 08544}}\\[26mm]

 {\large \bf{Abstract}}

 \end{center}

 \vspace{.2cm}

 \noindent The holographic principle is studied in the context of a
$n+1$ dimensional radiation dominated closed Friedman-Robertson-Walker
(FRW) universe. The radiation is represented by a conformal field
theory with a large central charge. Following recent ideas on
holography, it is argued that the entropy density in the early universe
is bounded by a multiple of the Hubble constant. The entropy of the CFT
is expressed in terms of the energy and the Casimir energy via a
universal Cardy formula that is valid for all dimensions. A new purely
holographic bound is postulated which restricts the sub-extensive
entropy associated with the Casimir energy. Unlike the Hubble bound,
the new bound remains valid throughout the cosmological evolution. When
the new bound is saturated the Friedman equation exactly coincides with
the universal Cardy formula, and the temperature is uniquely fixed in
terms of the Hubble parameter and its time-derivative. 

\end{titlepage}

 \newsection{Introduction}

 The holographic principle is based on the idea that for a given volume
$V$ the state of maximal entropy is given by the largest black hole
that fits inside $V$. 't Hooft and Susskind \cite{holog} argued on this
basis that the microscopic entropy $S$ associated with the volume $V$
should be less than the Bekenstein-Hawking entropy 
\be 
S\leq
{\frac{A}{4G}}   \label{bound} 
\ee 
of a black hole with horizon area
$A$ equal to the surface area of the boundary of $V$. Here the
dependence on Newton's constant $G$ is made explicit, but as usual
$\hbar$ and $c$ are set to one.

 To shed further light on the holographic principle and the entropy
bounds derived from it, we study in this paper the standard cosmology
of a closed radiation dominated Friedman-Robertson-Walker (FRW)
universe with general space-time dimension $$D=n+1.$$ The metric takes
the form 
\be 
ds^2=-dt^2+R^2(t)d\Omega^2_n 
\ee 
where $R(t)$ represents
the radius of the universe at a given time $t$ and $d\Omega^2_n$ is a
short hand notation for the metric on the unit $n$-sphere $S^n$. Hence,
the spatial section of a $(n\!+\!1)$d closed FRW universe is an
$n$-sphere with a finite volume 
$$ 
V={\rm Vol}(S^n)R^n. 
$$ 
The
holographic bound is in its naive form (\ref{bound}) not really
applicable to a closed universe, since space has no boundary.
Furthermore, the argumentation leading to (\ref{bound}) assumes that
it's possible to form a black hole that fills the entire volume. This
is not true in a cosmological setting, because the expansion rate $H$
of the universe as well as the given value of the total energy $E$
restrict the maximal size of black hole. As will be discussed in this
paper, this will lead to a modified version of the holographic bound.

 The radiation in an FRW universe is usually described by free or
weakly interacting mass-less particles.  More generally, however, one
can describe the radiation by an interacting conformal field theory
(CFT). The number of species of mass-less particles translates into
the value of the central charge $c$ of the CFT. In this paper we will
be particularly interested in radiation described by a CFT with a very
large central charge. In a finite volume the energy $E$ has a Casimir
contribution proportional to $c$. Due to this Casimir effect, the
entropy $S$ is no longer a purely extensive function of $E$ and $V$.
The entropy of a $(1\!+\!1)$d CFT is given by the well-known Cardy
formula \cite{cardy}
\be 
\label{Cardy} 
S=2\pi \sqrt{\frac{c}{6}\left(
L_{0}-\frac{c}{24}\right)}, 
\ee 
where $L_0$ represents the product $ER$ of the energy and radius, and the shift of $c\over 24$ is caused by the Casimir effect. In this paper we show that, after making the appropriate identifications for $L_0$ and $c$, the same Cardy formula is also valid for CFTs in other dimensions. This is rather surprising, since the standard derivation of the Cardy formula based on modular invariance only appears to work for $n=1$. By defining the central charge $c$ in terms of the Casimir energy, we are able to argue that the Cardy formula is universally valid. Specifically, we will show that with the appropriate identifications, the entropy $S$ for a $n\!+\!1$ dimensional CFT with an AdS-dual is exactly given by (\ref{Cardy}).

 The main new result of this paper is the appearance of a deep and
fundamental connection between the holographic principle, the entropy
formulas for the CFT, and the FRW equations for a radiation dominated
universe. In $n\!+\!1$ dimensions the FRW equations are given by 
\ba
\label{friedman1} H^2\is {16\pi G\over n(n\!-\!1)}{E\over V}-{1\over
R^2}\\[1mm] 
\label {friedman2} \dot{H}\is -\frac{8\pi
G}{n\!-\!1}\left({E\over V} +p\right) +{1\over R^2} 
\ea 
where
$H=\dot{R}/R$ is the Hubble parameter, and the dot denotes as usual
differentiation with respect to the time $t$. The FRW equations are
usually written in terms of the energy density $\rho=E/V$, but for the
present study it is more convenient to work with the total energy $E$
and entropy $S$ instead of their respective densities $\rho$ and
$s=S/V$. Note that the cosmological constant has been put to zero; the
case $\Lambda\neq 0$ will be described elsewhere \cite{ivo}.

 Entropy and energy momentum conservation together with the equation of
state $p=E/nV$ imply that $E/V$ and $p$ decrease in the usual way like
$R^{-(n+1)}$.  Hence, the cosmological evolution follows the standard
scenario for a closed radiation dominated FRW universe. After the
initial Big Bang, the universe expands until it reaches a maximum
radius, the universe subsequently re-collapses and ends with a Big
Crunch. No surprises happen in this respect.

 The fun starts when one compares the holographic entropy bound with
the entropy formulas for the CFT. We will show that when the 
bound is saturated the FRW equations and entropy formulas of the CFT merge together into one set of equation.  One easily checks on the back of an envelope that via the substitutions 
\ba 
2\pi L_0\ & \Rightarrow &\  {2\pi \over n}ER\nonumber\\[1mm]
2\pi {c\over 12}\ & \Rightarrow &\  (n\!-\!1){V\over 4GR}\label{match} \\[1mm] \nonumber S\ & \Rightarrow &\  (n\!-\!1){HV\over 4G}
\ea 
the Cardy formula (\ref{Cardy}) {\it exactly} turns into the $n+1$ dimensional Friedman equation (\ref{friedman1}). This observation appears as a natural consequence of the holographic principle. In sections 2 and 3 we introduce  three cosmological bounds each corresponding to one of the equations in (\ref{match})  The Cardy formula is presented and derived in section 4. In section 5 we introduce a new cosmological bound, and show that the FRW equations and the entropy formulas are exactly matched when the bound is saturated. In section 6 we present a graphical picture of the entropy bounds and their time evolution.

\bigskip

\newsection{Cosmological entropy bounds}

 This section is devoted to the description of three cosmological
entropy bounds: the Bekenstein bound, the holographic
Bekenstein-Hawking bound, and the Hubble bound. The relation with the
holographic bound proposed by Fischler-Susskind and Bousso (FSB) will
also be clarified.

 \medskip

 \itsubsection{The Bekenstein bound}

 Bekenstein \cite{bekenstein} was the first to propose a bound
on the entropy of a macroscopic system. He argued that for a system with
limited self-gravity, the total entropy $S$ is less or equal than a
multiple of the product of the energy and the linear size of the
system. In the present context, namely that of a closed radiation
dominated FRW universe with radius $R$, the appropriately normalized
Bekenstein bound is \be S\leq S_B \ee where the Bekenstein entropy
$S_B$ is defined by \be \label{SB} S_B\,\equiv\, {2\pi\over n} ER. \ee
The bound is most powerful for relatively low energy density or small
volumes. This is due to the fact that $S_B$ is super-extensive: under
$V\to\lambda V$ and $E\to \lambda E$ it scales like $S_B\to
\lambda^{1+1/n}S_B$.

 For a radiation dominated universe the Bekenstein entropy is constant
throughout the entire evolution, since $E\sim R^{-1}$.  Therefore, once
the Bekenstein bound is satisfied at one instance, it will remain
satisfied at all times as long as the entropy $S$ does not change. The Bekenstein entropy is the most natural generalization of the Virasoro operator $2\pi L_0$ to arbitrary dimensions, as is apparent from (\ref{match}). Indeed, it is useful to think about $S_B$ not really as an entropy but rather as the energy measured with respect to an appropriately chosen conformal time coordinate. 

\medskip

\itsubsection{The Bekenstein-Hawking bound}

 The Bekenstein-bound is supposed to hold for systems with limited
self-gravity, which means that the gravitational self-energy of the
system is small compared to the total energy $E$. In the current
situation this implies, concretely, that the Hubble radius $H^{-1}$ is
larger than the radius $R$ of the universe. So the Bekenstein bound is
only appropriate in the parameter range $HR\leq 1$. In a strongly
self-gravitating universe, that is for $HR\geq 1$, the possibility of
black hole formation has to be taken into account, and the entropy
bound must be modified accordingly. Here the general philosophy of the
holographic principle becomes important.

 It follows directly from the Friedman equation (\ref{friedman1}) that
\be 
HR \begin{array}{c}\ {}_< \\[-1.5mm] \ {}^>\end{array}
1\qquad\Leftrightarrow\qquad S_B \begin{array}{c}\ {}_< \\[-1.5mm] \
{}^>\end{array} (n\!-\!1){V\over 4G R} 
\ee 
Therefore, to decide whether
a system is strongly or weakly gravitating one should compare the
Bekenstein entropy $S_B$ with the quantity \be \label{SBH} S_{BH}\,
\equiv\, (n\!-\!1){V\over 4GR}. \ee When $S_B\leq S_{BH}$ the system is
weakly gravitating, while for $S_B\geq S_{BH}$ the self-gravity is
strong. We will identify $S_{BH}$ with the holographic
Bekenstein-Hawking entropy of a black hole with the size of the
universe. $S_{BH}$ indeed grows like an area instead of the volume, and
for a closed universe it is the closest one can come to the usual
expression $A/4G$.

 As will become clear in this paper, the role of $S_{BH}$ is not to
serve as a bound on the total entropy, but rather on a sub-extensive
component of the entropy that is association with the Casimir energy of
the CFT. The relation (\ref{match}) suggests that the Bekenstein-Hawking entropy is closely related to the central charge $c$. Indeed, it is well-known from $(1\!+\!1)d$ CFT that the central charge characterizes the number of degrees of freedom may be even better than the entropy. This fact will be further explained in sections 5 and 6, when we describe a new cosmological bound on the Casimir energy and its associated entropy.

 \medskip

 \itsubsection{The Hubble entropy bound}

 The Bekenstein entropy $S_B$ is equal to the holographic
Bekenstein-Hawking entropy $S_{BH}$ precisely when $HR=1$. For $HR>1$
one has $S_B>S_{BH}$ and the Bekenstein bound has to be replaced by a
holographic bound. A naive application of the holographic principle
would imply that the total entropy $S$ should be bounded by $S_{BH}$.
This turns out to be incorrect, however, since a purely holographic
bound assumes the existence of arbitrarily large black holes, and is
irreconcilable with a finite homogeneous entropy density.

 Following earlier work by Fischler and Susskind \cite{FS}, it was
argued by  Easther and Lowe \cite{lowe}, Veneziano \cite{GV}, Bak and Rey \cite{rey}, Kaloper and Linde \cite{linde}, that the maximal entropy inside the universe is produced by black holes of the size of the Hubble horizon, see also \cite{brustein}. Following the usual holographic arguments
one then finds that the total entropy should be less or equal than the
Bekenstein-Hawking entropy of a Hubble size black hole times the number
$N_H$ of Hubble regions in the universe. The entropy of a Hubble size
black hole is roughly $HV_H/4G$, where $V_H$ is the volume of a single
Hubble region. Combined with the fact that $N_H=V/V_H$ one obtains an
upper bound on the total entropy $S$ given by a multiple of $HV/4G$.
The presented arguments of \cite{lowe,rey,linde,GV} are not sufficient
to determine the precise pre-factor, but in the following subsection we
will fix the normalization of the bound by using a local version of the
Fischler-Susskind-Bousso formulation of the holographic principle. The
appropriately normalized entropy bound takes the form 
\be
\label{hubblebound} S\leq S_H \qquad\quad \mbox{for $\qquad HR\geq 1$}
\ee 
with 
\be 
\label{SH} S_{H}\equiv(n\!-\!1){HV\over 4G}. 
\ee 
The Hubble bound is only valid for $HR\geq 1$. In fact, it is easily seen
that for $HR\leq 1$ the bound will at some point be violated. For
example, when the universe reaches its maximum radius and starts to
re-collapse the Hubble constant $H$ vanishes, while the entropy is
still non-zero.\footnote{To avoid this problem a different covariant version of the Hubble bound was proposed in \cite{bruven}.} This should not really come as a surprise, since the
Hubble bound was based on the idea that the maximum size of a black
hole is equal to the Hubble radius. Clearly, when the radius $R$ of the
universe is smaller than the Hubble radius $H^{-1}$ one should
reconsider the validity of the bound. In this situation, the
self-gravity of the universe is less important, and the appropriate
entropy bound is 
\be 
S \leq S_B \qquad\quad
\mbox{for $\qquad HR\leq 1$} 
\ee

\itsubsection{The Hubble bound and the FSB prescription.}

Fischler, Susskind, and subsequently Bousso \cite{bousso}, have proposed an ingenious
version of the holographic bound that restricts the entropy flow
through contracting light sheets. The FSB-bound works well in many
situations, but, so far, no microscopic derivation has been given. Wald
and collaborators \cite{wald} have shown that the FSB bound follows
from local inequalities on the entropy density and the stress energy.
The analysis of \cite{wald} suggests the existence a local version of
the FSB entropy bound, one that does not involve global information
about the causal structure of the universe, see also \cite{bruven}. The idea of to formulate the holographic principle via entropy flow through light sheets also occurred in the work of Jacobson \cite{jacobson}, who used it to derive an intriguing relation between the Einstein equations and the first law of thermodynamics. In this subsection, a local FSB bound will be presented that leads to a precisely normalized upper limit on the entropy in terms of the Hubble constant.

 According to the original FSB proposal, the entropy flow $S$ through a
contracting light sheet is less or equal to $A/4G$, where $A$ is the
area of the surface from which the light sheet originates. The following infinitesimal version of this FSB prescription will lead to the Hubble bound.
For every $n\!-\!1$ dimensional surface at time $t+dt$ with area $A+dA$
one demands that 
\be 
\label{fsb} 
dS\leq {dA\over 4G}, 
\ee 
where $dS$
denotes the entropy flow through the infinitesimal light sheets
originating at the surface at $t+dt$ and extending back to time $t$,
and $dA$ represents the increase in area between $t$ and $t+dt$. For a
surface that is kept fixed in co-moving coordinates the area $A$
changes as a result of the Hubble expansion by an amount 
\be 
\label{dA}
dA=(n\!-\!1) H A\,dt, 
\ee 
where the factor $n\!-\!1$ simply follows
from the fact that $A\sim R^{n-1}$. Now pick one of the two past
light-sheets that originate at the surface: the inward or the outward
going. The entropy flow through this light-sheet between $t$ and $t+dt$
is given by the entropy density $s=S/V$ times the infinitesimal volume
$A dt$ swept out by the light-sheet. Hence, 
\be 
\label{dS} dS={S\over V} A\, 
dt. 
\ee 
By inserting this result together with (\ref{dA}) into
the infinitesimal FSB bound (\ref{fsb}) one finds that the factor $Adt$
cancels on both sides and one is left exactly with the Hubble bound
$S\leq S_H$ with the Hubble entropy $S_H$ given in (\ref{SH}). We
stress that the relation with the FSB bound was merely used to fix the
normalization of the Hubble bound, and should not be seen as a
derivation.
 \bigskip

\newsection{Time-evolution of the entropy bounds.}

Let us now return to the three cosmological entropy bounds discussed in section 2. The Friedman equation (\ref{friedman1}) can be re-written as an
identity that relates the Bekenstein-, the Hubble-, and the
Bekenstein-Hawking entropy. One easily verifies that the expressions
given in (\ref{SB}), (\ref{SBH}), and (\ref{SH}) satisfy the quadratic
relation 
\be 
\label{SSC} S_H^2+(S_{B}-S_{BH})^2=S_B^2. 
\ee  
It is deliberately written in a Pythagorean form, since it suggests 
a useful graphical picture of the three entropy bounds. 
By representing each entropy by a line with length equal to its value one
finds that due to the quadratic Friedman relation (\ref{SSC}) all three
fit nicely together in one diagram, see figure 1. The circular form of
the diagram reflects the fact that $S_B$ is constant during the
cosmological evolution. Only $S_H$ and $S_{BH}$ depend on time.

 Let us introduce a conformal time coordinate via 
\be R 
d\eta=(n\!-\!1) dt 
\ee 
and let us compute the $\eta$-dependence of $S_{BH}$
and $S_H$. For $S_{BH}$ this easily follows from:
$\dot{S}_{BH}=(n\!-\!1) H S_{BH}= (n\!-\!1) R^{-1}S_{H}$. For $S_H$ the
calculation is a bit more tedious, but with the help of the FRW
equations, the result can eventually be put in the form 
\ba
\label{rotate} {dS_H\over d\eta} \is  S_B-S_{BH},\nonumber \\
{dS_{BH}\over d\eta}\is \ -S_H. 
\ea 
These equations show that the
conformal time coordinate $\eta$ can be identified with the angle
$\eta$, as already indicated in figure 1. As time evolves the Hubble
entropy $S_H$ rotates into the combination $S_B-S_{BH}$ and visa
versa. Equation (\ref{rotate}) can be integrated to 
\ba 
\label{defth}
S_H\is \ S_B \sin\eta\nonumber\\ S_{BH}\is
S_B(1-\cos\eta)\qquad{}\mbox{$ $} 
\ea 
The conformal time coordinate
$\eta$ plays the role of the time on a cosmological clock that only
goes around once: at $\eta=0$ time starts with a Big Bang and at
$\eta=2\pi$ it ends with a Big Crunch. Note that $\eta$ is related to
the parameter $HR$ via 
\be 
HR=\cot{\eta\over 2} 
\ee 
\figuur{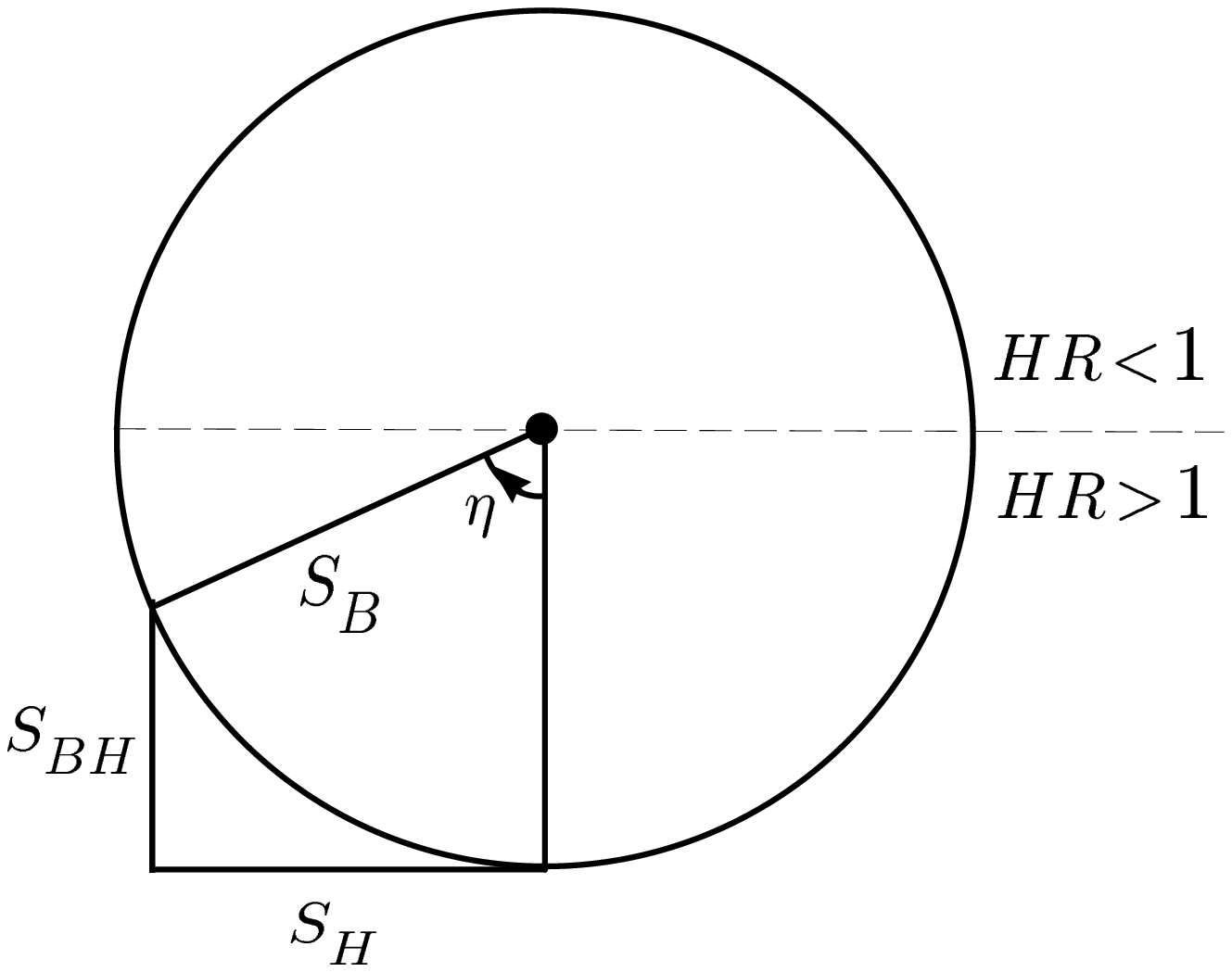}{7cm}{A graphical representation of the Bekenstein
entropy $S_B$, the Hubble entropy $S_H$ and the Bekenstein-Hawking
entropy $S_{BH}$. The angle $\eta$
corresponds to the conformal time coordinate. The value of each entropy
is represented by an actual distance: $S_B$  is constant, while $S_H$
and $S_{BH}$ change with time.}
So far we have not
yet included the CFT into our discussion. We will see that the entropy
of the CFT will `fill' part of the diagram, and in this way give rise
to a special moment in time when the entropy bounds are saturated.
\nopagebreak

\newpage

\newsection{Casimir energy and the Cardy formula}

We now turn to the discussion of the entropy of the CFT that lives
inside the FRW universe.  We begin with a study of the finite
temperature Casimir energy with the aim to exhibit its relation with
the entropy of the CFT. Subsequently a universal Cardy formula will be
derived that expresses the entropy in terms of the energy and the
Casimir energy, and is valid for all values of the spatial dimension
$n$.

 \itsubsection{The Euler relation and Casimir energy.}

 In standard textbooks on cosmology \cite{weinberg,kolbturner} it is
usually assumed that the total entropy $S$ and energy $E$ are extensive
quantities. This fact is used for example to relate the entropy density
$s$ to the energy density $\rho$ and pressure $p$, via $Ts=\rho+p$. For
a thermodynamic system in finite volume $V$ the energy $E(S,V)$,
regarded as a function of entropy and volume,  is called extensive when
it satisfies $E(\lambda S,\lambda V)=\lambda E(S,V).$ Differentiating
with respect to $\lambda$ and putting $\lambda=1$ leads to the Euler
relation\footnote{We assume here that there are no other thermodynamic
functions like a chemical or electric potential. For a system with a
1st law like $TdS=dE+pdV+\mu dN+\Phi dQ$ the Euler relation reads
$TS=E+pV+\mu N+ \Phi Q$.} \be \label{ext} E= V\left(\frac{\partial
E}{\partial V}\right)_S +S\left(\frac{\partial E}{\partial S}\right)_V
\ee The first law of thermodynamics $dE=TdS-pdV$ can now be used to
re-express the derivatives via the thermodynamic relations \be
\label{ident} \left(\frac{\partial E}{\partial V}\right)_S=-p,\qquad
\left(\frac{\partial E}{\partial S}\right)_V=T. \ee The resulting
equation $TS=E+pV$ is equivalent to the previously mentioned relation
for the entropy density $s$.

 For a CFT with a large central charge the entropy and energy are not
purely extensive. In a finite volume the energy $E$ of a CFT contains a
non-extensive Casimir contribution proportional to $c$. This is well
known in $(1\!+\!1)$ dimensions where it gives rise to the familiar
shift of $c/24$ in the $L_0$ Virasoro operator. The Casimir energy
is the result of finite size effects in the quantum
fluctuations of the CFT, and disappears when the volume becomes
infinitely large. It therefore leads to sub-extensive contributions to
the total energy $E$. Usually the Casimir effect is discussed at zero
temperature  \cite{casimir}, but a similar effect occurs at finite temperature. The
value of the Casimir energy will in that case generically depend on the
temperature $T$.

 We will now define the Casimir energy as the violation of the Euler
identity (\ref{ext}) \be \label{ECdef} E_C\equiv n(E+pV-TS) \ee Here we
inserted for convenience a factor equal to the spatial dimension $n$.
{}From the previous discussion it is clear that $E_C$ parameterizes the
sub-extensive part of the total energy. The Casimir energy will just as
the total energy be a function of the entropy $S$ and the volume $V$.
Under $S\to\lambda S$ and $V\to \lambda V$ it scales with a power of
$\lambda$ that is smaller than one. On general grounds one expects that
the first subleading correction to the extensive part of the energy
scales like 
\be 
E_C(\lambda S,\lambda V)=\lambda^{1-2/n}E_C(S,V) 
\ee
One possible way to see this is to write the energy as an integral over
a local density expressed in the metric and its derivatives.
Derivatives scale like $\lambda^{-1/n}$ and because derivatives come
generally in pairs, the first subleading terms indeed has two
additional factors of $\lambda^{-1/n}$. \medskip The total energy
$E(S,V)$ may be written as a sum of two terms 
\be \label{EEE}
E(S,V)=E_E(S,V)+{1\over 2} E_C(S,V) 
\ee where the first term $E_E$
denotes the purely extensive part of the energy $E$ and $E_C$
represents the Casimir energy. Again the factor $1/2$ has been put in
for later convenience. By repeating the steps that lead to the Euler
relation one easily verifies the defining equation (\ref{ECdef}) for
the Casimir energy $E_C$.

\itsubsection{Universality of the Cardy formula and the Bekenstein
bound}

 Conformal invariance implies that the product $ER$ is independent of
the volume $V$, and is only a function of the entropy $S$. This holds
for both terms $E_E$ and $E_C$ in (\ref{EEE}). Combined with the known
(sub-)extensive behavior of $E_E$ and $E_C$ this leads to the following
general expressions $$ E_E={a\over 4\pi R}S^{1+{1/n}}\qquad\qquad
E_C={b\over 2\pi R} S^{1-{1/n}} $$ where $a$ and $b$ are a priori
arbitrary positive coefficients, independent of $R$ and $S$. The
factors of $4\pi$ and $2\pi$ are put in for convenience. With these
expressions, one now easily checks that the entropy $S$ can be written
as \be \label{GCab} S={2\pi R \over \sqrt{ab}}\sqrt{E_C(2E-E_C)}. \ee
If we ignore for a moment the normalization, this is exactly the Cardy
formula: insert $ER=L_0$ and $E_CR=c/12$, and one recovers
(\ref{Cardy}). It is obviously an interesting question to compute the
coefficients $a$ and $b$ for various known conformal invariant field
theories. This should be particularly straightforward for free field
theories, such as $d=4$ Maxwell theory and the self-dual tensor theory
in $d=6$. This question is left for future study.

 Given the energy $E$ the expression (\ref{GCab}) has a maximum value.
For all values of $E$, $E_C$ and $R$ one has the inequality $$ S
\leq{\, 2\pi\over \sqrt{ab}}ER $$ This looks exactly like the
Bekenstein bound, except that the pre-factor is in general different
from the factor $2\pi/n$ used in the previous section. In fact, in the
following subsection we will show that for CFTs with an AdS-dual
description, the value of the product $ab$ is exactly equal to $n^2$,
so the upper limit is indeed exactly given by the Bekenstein entropy.
Although we have no proof of this fact, we believe that the Bekenstein
bound is universal. This implies that the product $ab$ for all CFTs in
$n\!+\!1$ dimensions is larger or equal than $n^2$. Only then it is
guaranteed that the upper limit on the entropy is less or equal than
$S_B$.

 The upper limit is reached when the Casimir energy $E_C$ is equal to
the total energy $E$. Formally, when $E_C$ becomes larger that $E$ the
entropy $S$ will again decrease. Although in principle this is
possible, we believe that in actual examples the Casimir energy $E_C$
is bounded by the total energy $E$. So, from now on we assume that \be
E_C\leq E \ee In the next subsection we provide further evidence for
this inequality.

 {}From now on we will assume that we are dealing with a CFT for which
$ab=n^2$. In the next section I will show that this includes all CFTs
that have an AdS-dual description.

\itsubsection{The Cardy formula derived from AdS/CFT}

 Soon after Maldacena's AdS/CFT-correspondence \cite{malda} was
properly understood \cite{GKP,W} it was convincingly argued by Witten
\cite{witten} that the entropy, energy and temperature of CFT at high
temperatures can be identified with the entropy, mass, and Hawking
temperature of the AdS black hole previously considered by Hawking and
Page \cite{hawpage}. Using this duality relation the following
expressions can be derived for the energy and entropy\footnote{These
expressions differ somewhat from the presented formulas in
\cite{witten} due to the fact that (i) the $D+1$ dimensional Newton
constant has been eliminated using its relation with the central
charge, (ii) the coordinates have been re-scaled so that the CFT lives
on a sphere with radius equal to the black hole horizon. We will not
discuss the AdS perspective in this paper, since the essential physics
can be understood without introducing an extra dimension. The
discussion of the CFT/FRW cosmology from an AdS perspective will be
described elsewhere \cite{ivo}.} for a $D=n+1$ dimensional CFT on
$R\times S^{n}$: 
\ba 
S\is \frac{c}{12}\frac{V}{L^{n}} \nonumber \\ 
E\is \frac{c}{12}\frac{n}{4\pi L}\left(1+\frac{L^{2}}{R^{2}}\right)
\frac{V}{L^n} \label{SE} 
\ea 
The temperature again follows from the
first law of thermodynamics. One finds 
\be \label{T} T=\frac{1}{4\pi
L}\left( \left( n+1\right) +\left( n-1\right) \frac{L^{2}}{R^{2}}\right). 
\ee 
The length scale $L$ of the thermal CFT
arises in the AdS/CFT correspondence as the curvature radius of the AdS
black hole geometry. The expression for the energy clearly exhibits a
non-extensive contribution, while also the temperature $T$ contains a
corresponding non-intensive term. Inserting the equations
(\ref{SE},\ref{T}) into (\ref{ECdef}) yields the following result for
the Casimir energy 
\be 
\label{ECC} 
E_C= \frac{c}{12}{n\over 2\pi R}\frac{V}{L^{n-1}R}. 
\ee 
Now let us come to the Cardy formula. The
entropy $S$, energy $E$ and Casimir energy $E_C$ are expressed in $c$,
$L$ and $R$. Eliminating $c$ and $L$ leads to a unique expression for
$S$ in terms of $E$, $E_C$ and $R$. One easily checks that it takes the
form of the Cardy formula 
\be 
\label{GC} 
S={2\pi R\over n}\sqrt{E_C\left(2E-E_C \right)} 
\ee 
In the derivation of these formulas it was assumed that $R>\!\!>L$. 
One may worry therefore that these formulas are not applicable in the early universe. Fortunately this is not a problem because during an adiabatic expansion both $L$ and $R$ scale in the same way so that $R/L$ is fixed. Hence the formulas are valid provided the (fixed) ratio of the thermal
wave-length and the radius $R$ is much smaller than one. Effectively
this means, as far as the CFT is concerned, we are in a high
temperature regime. We note further that with in this parameter range,
the Casimir energy $E_C$ is indeed smaller than the total energy $E$.

 Henceforth, we will assume that the CFT that describes the radiation
in the FRW universe will have an entropy given by (\ref{GC}) with the
specific normalization of $2\pi/n$. Note that if we take $n=1$ and make
the previously mentioned identifications $ER=L_0$ and $E_CR=c/12$ that
this equation exactly coincides with the usual Cardy formula. We will
therefore in the following refer to (\ref{GC}) simply as the Cardy
formula. To check the precise coefficient of the Cardy formula for a
CFT we have made use of the AdS/CFT correspondence. The rest of our
discussions in the preceding and in the following sections do not
depend on this correspondence. So, in this paper we will not make use
of any additional dimensions other than the ones present in the
FRW-universe.

\bigskip

\newsection{A new cosmological bound}

In this section a new cosmological bound will be presented, which is equivalent to the Hubble bound in the strongly gravitating phase, but which unlike the Hubble bound remains valid in the phase of weak self-gravity. When the bound is saturated the FRW equations and the CFT formulas for the entropy and Casimir energy completely coincide.

 \medskip

\itsubsection{A cosmological bound on the Casimir energy}

 Let us begin by presenting another criterion for distinguishing
between a weakly or strongly self-gravitating universe. When the
universe goes from the strongly to the weakly self-gravitating phase,
or vice-versa, the Bekenstein entropy $S_B$ and the Bekenstein-Hawking
entropy $S_{BH}$ are equal in value. Given the radius $R$, we now
define the `Bekenstein-Hawking' energy $E_{BH}$ as the value of the
energy $E$ for which $S_B$ and $S_{BH}$ are exactly equal. This leads
to the condition 
\be 
\label{EBH} 
{2\pi\over
n}E_{BH}R\equiv(n\!-\!1){V\over 4GR}. 
\ee 
One may interpret $E_{BH}$ as
the energy required to form a black hole with the size of the entire
universe. Now, one easily verifies that \ba \label{weakstrong} E & \leq
& E_{BH} \qquad \mbox{for $\qquad HR\leq 1$}\nonumber\\ E & \geq &
E_{BH} \qquad \mbox{for $\qquad HR\geq 1$}. \ea Hence, the universe is
weakly self-gravitating when the total energy $E$ is less than $E_{BH}$
and strongly gravitating for $E>E_{BH}$.

 We are now ready to present a proposal for a new cosmological bound. 
It is not formulated as a bound on the entropy $S$, but as a restriction on the
Casimir energy $E_C$. 
The physical content of the bound is the Casimir energy $E_C$ by
itself can not be sufficient to form a universe-size black hole.
Concretely, this implies that the Casimir energy $E_C$ is less or equal
to the Bekenstein-Hawking energy $E_{BH}$. Hence, we postulate 
\be
E_C\leq E_{BH} 
\ee 
To put the bound in a more conventional notation one
may insert the definition (\ref{ECdef}) of the Casimir energy together
with the defining relation (\ref{EBH}) of the Bekenstein-Hawking
energy. We leave this to the reader.

The virtues of the new cosmological bound are: (i) it
is universally valid and does not break down for a weakly gravitating
universe, (ii) in a strongly gravitating universe it is equivalent to
the Hubble bound, (iii) it is purely holographic and can be formulated
in terms of the Bekenstein-Hawking entropy $S_{BH}$ of a universe-size
black hole, (iv) when the bound is saturated the laws of general
relativity and quantum field theory converge in a miraculous way,
giving a strong indication that they have a common origin in a more
fundamental unified theory. 

The first point on the list is easily checked because $E_C$ decays like $R^{-1}$ while $E_{BH}$ goes like $R^{-n}$. Only when the universe
re-collapses and returns to the strongly gravitating phase the bound may
again become saturated. To be able to proof the other points on the list of advertised virtues, we have to take a closer look to the FRW equations and the CFT formulas for the entropy an entropy. 

\medskip

\itsubsection{A cosmological Cardy formula}

To show the equivalence of the new bound with the Hubble bound 
let us write the Friedman equation as an expression for the Hubble
entropy $S_H$ in terms of the energy $E$, the radius $R$ and the
Bekenstein-Hawking energy $E_{BH}$. Here, the latter is used to remove
the explicit dependence on Newton's constant $G$. The resulting
expression is unique and takes the form 
\be 
\label{CCF} S_H={2\pi\over n}R\sqrt{E_{BH}\left(2E-E_{BH}\right)} 
\ee 
This is exactly the Cardy formula (\ref{GC}), except that the role of the Casimir energy $E_C$ in CFT formula is now replaced by the Bekenstein-Hawking
energy $E_{BH}$. Somehow, miraculously, the Friedman equation knows about the Cardy formula for the entropy of a CFT!

With the help of (\ref{CCF}) is now a straightforward matter to proof that when $HR\geq 1$ the new bound $E_C\leq E_{BH}$ is
equivalent to the Hubble bound $S\leq S_H$.  
First, let us remind that for $HR\geq 1$ the energy $E$ satisfies  $E\geq E_{BH}$. Furthermore, we always assume that the Casimir energy $E_C$ is smaller than the total energy $E$. The entropy $S$ is a monotonically increasing function of $E_C$ as long as $E_C\leq E$. Therefore in the range 
\be
\label{range}
E_C\leq E_{BH}\leq E
\ee
the maximum entropy is reached when $E_C=E_{BH}$. In
that case the Cardy formula (\ref{GC}) for $S$ exactly turns into the
cosmological Cardy formula (\ref{CCF}) for $S_H$. Therefore, we conclude that $S_H$  is indeed the maximum entropy that can be reached when $HR\geq 1$. Note that in the weakly self-gravitating phase, when $E\leq E_{BH}$, the maximum is reached earlier, namely for $E_C=E$. The maximum entropy is in that case given by the bekenstein entropy $S_B$. The bifurcation of the new bound in two entropy bounds is a direct consequence of the fact that the Hubble bound is written as the square-root of a quadratic expression.

\medskip

\itsubsection{A limiting temperature}

So far we have focussed on the entropy and energy of the CFT and on
the first of the two FRW equations, usually referred to as the Friedman
equation. We will now show that also the second FRW equation has a
counterpart in the CFT, and will lead to a constraint on the temperature $T$. 
Specifically, we will find that the bound on $E_C$ implies that the temperature $T$ in the early universe is bounded from below by
\be
T_H\equiv-{\dot{H}\over 2\pi H}
\ee
The minus sign is necessary to get a positive result, since in a radiation dominated universe the expansion always slows down. Further, we assume that we are in the strongly self-gravitating phase with $HR\geq 1$, so that there is no danger of dividing by zero.

The second FRW equation in (\ref{friedman2}) can now be written as a relation between $E_{BH}$, $S_H$ and $T_H$ that takes the familiar form  
\be
E_{BH}= n(E+pV-T_H S_H)
\ee
This equation has exactly the same form as the defining relation $E_C=n(E+pV-TS)$ for the Casimir energy.
In the strongly gravitating phase we have just argued that the bound $E_C\leq E_{BH}$ is equivalent to the Hubble bound $S\leq S_H$. It follows immediately that the temperature $T$ in this phase is bounded from below by $T_H$. One has
\be
T\geq T_H\qquad\qquad{\mbox{for $\ HR\geq 1$}}
\ee 
When the cosmological bound is saturated all inequalities turn into equalities. 
The Cardy formula and the defining Euler relation for the Casimir energy in that case exactly match the Friedman equation for the Hubble constant and the FRW equation for its time derivative.

\figuren{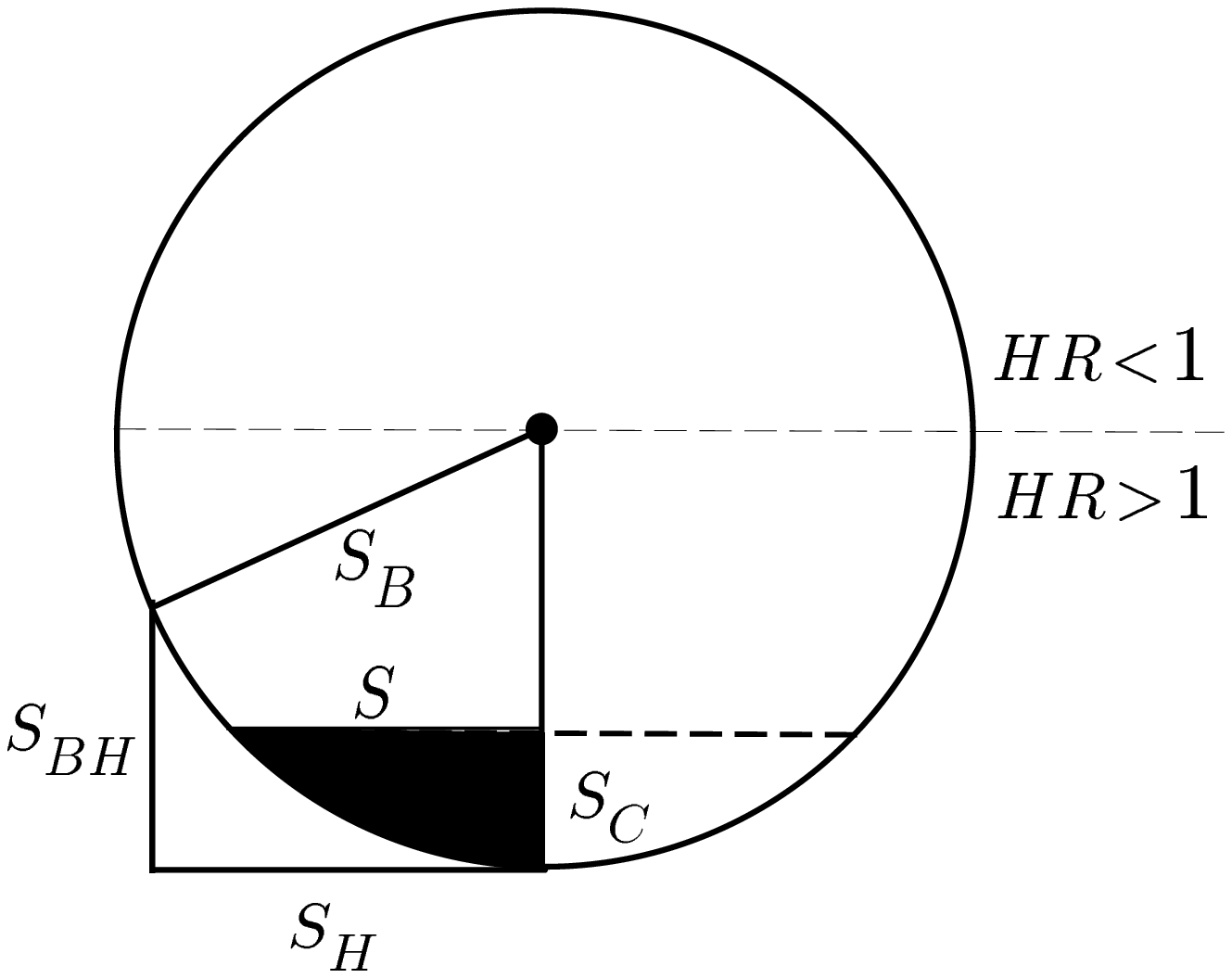}{7cm}{The entropy $S$ and Casimir entropy $S_C$ 
fill part of the cosmological entropy diagram. The diagram shows: (i) the Bekenstein bound $S\leq S_B$ is valid at all times (ii) the Hubble
bound $S\leq S_H$ restricts the allowed range of $\eta$ in the range $HR>1$, but is violated for $HR<1$, (iii) the new bound $S_C\leq S_{BH}$ is equivalent to the Hubble bound for $HR>1$, and remains valid for $HR<1$.}

\newsection{The entropy bounds revisited.}

We now return to the cosmological entropy bounds introduced in sections 2 and 3.
In particular, we are interested in the way that the entropy of the CFT may be incorporated in the entropy diagram described in section $3$.  For this purpose it will be useful to introduce a non-extensive component of the entropy that is associated with the Casimir energy.

The cosmological bound $E_C\leq E_{BH}$ can also be formulated 
as an entropy bound, not on the total entropy, but on a non-extensive
part of the entropy that is associated with the Casimir energy.
 In analogy with the definition of the Bekenstein entropy (\ref{SB})
one can introduce a 'Casimir' entropy defined by 
\be 
S_C\equiv
{2\pi\over n}E_C R. 
\ee 
For $d=(1\!+\!1)$ the Casimir entropy is
directly related to the central charge $c$. One has $S_C=2\pi\, c/12$.
In fact, it is more appropriate to interpret the Casimir entropy $S_C$
as a generalization of the central charge to $n\!+\!1$ dimensions than what
is usually called the central charge $c$. Indeed, if one introduces a
dimensionless `Virasoro operator' $\widetilde{L}_0\equiv {1\over
2\pi}S_B$ and a new central charge ${\widetilde{c}\over 12}\equiv
{1\over 2\pi}S_C$, the $n\!+\!1$ dimensional entropy formula (\ref{GC}) is
exactly identical to (\ref{Cardy}).

The Casimir entropy $S_C$ is sub-extensive because under $V\to\lambda
V$ and $E\to \lambda E$ it goes like $S_C\to \lambda^{1-1/n}S_C$. In
fact, it scales like an area! This is a clear indication that
the Casimir entropy has something to do with holography. The total
entropy $S$ contains extensive as well as sub-extensive contributions.
One can show that for $E_C\leq E$ the entropy $S$ satisfies the
following inequalities \be S_C\leq S\leq S_B \ee where both equal signs
can only hold simultaneously. The precise relation between $S$ and its
super- and sub-extensive counterparts $S_B$ and $S_C$ is determined by
the Cardy formula, which can be expressed as 
\be 
\label{SSS}
S^2+(S_B-S_C)^2=S_B^2. 
\ee 
This identity has exactly the same form as the
relation (\ref{SSC}) between the cosmological entropy bounds, except
that in (\ref{SSC}) the role of the entropy and Casimir entropy are
taken over by the Hubble entropy $S_H$ and Bekenstein-Hawking entropy
$S_{BH}$. This fact will be used to incorporate the entropy $S$ and the Casimir entropy $S_C$ in the entropy diagram introduce in section 3.

The cosmological bound on the Casimir energy presented in the section
4 can be formulated as an upper limit on the Casimir entropy $S_C$.
{}From the definitions of $S_C$ and $E_{BH}$ it follows directly that
the bound $E_C\leq E_{BH}$ is equivalent to \be S_C\leq S_{BH} \ee
where we made use of the relation (\ref{EBH}) to re-write $E_{BH}$
again in terms of the Bekenstein-Hawking entropy $S_{BH}$.  Thus the
bound puts a holographic upper limit on  the d.o.f. of the CFT as
measured by the Casimir entropy $S_C$.

In figure 2 we have graphically depicted the quadratic relation
between the total entropy $S$ and the Casimir entropy $S_C$ in the same
diagram we used to related the cosmological entropy bounds. From this
diagram it easy to determine the relation between the new bound and the
Hubble bound. One clearly sees that when $HR> 1$ that the two bounds
are in fact equivalent. When the new bound is saturated, which means
$S_C=S_{BH}$, then the Hubble bound is also saturated, {\it ie.}
$S=S_H$. The converse is not true: there are two moments in the region
$HR<1$ when the $S=S_H$, but $S_C\neq S_{BH}$. In our opinion, this is
an indication that the bound on the Casimir energy has a good chance of
being  a truly fundamental bound.

\bigskip

\newsection{Summary and conclusion}

In this paper we have used the holographic principle to study the bounds on the entropy in a radiation dominated universe. The radiation has been described by a continuum CFT in the bulk. Surprisingly the CFT appears to know about the holographic entropy bounds, and equally
surprising the FRW-equations know about the entropy formulas for the
CFT. Our main results are summarized in the following two tables. Table 1. contains an overview of the bounds that hold in the early universe on the temperature, entropy and Casimir energy. In table 2. the Cardy formula for the CFT and the Euler relation for the Casimir energy are matched with the Friedman equations written in terms of the quantities listed in table 1. 
\medskip

\begin{center}
\begin{tabular}{||c|c||}
\hline\hline 
{\huge $\mbox{\normalsize{\em CFT-bound}}$} & {\em FRW-definition}{\Huge ${}$}\\
\hline\hline
& 
\\
$\ T\geq T_H\  $& $T_H\equiv-{\dot{H}/2\pi H}$\\ 
& 
\\
\hline
& 
\\
$\ S\leq S_H\  $ & $S_H \equiv {(n\!-\!1) HV/4G}$ \\
& 
\\
\hline
& 
\\
$\ E_C\leq E_{BH}\ $ & $\ E_{BH}\equiv {n(n\!-\!1)V/8\pi GR^2}\ $   
\\
&
\\
\hline\hline
\end{tabular}

\medskip

{\it Table 1: summary of cosmological bounds}
\end{center}

\newpage

${}$
\vspace{-.8cm}

\begin{center}

\begin{tabular}{||c|c||}
\hline \hline 
{\em CFT-formula} & {\em FRW-equation}
\\
\hline\hline
& 
\\
$\ S={2\pi R\over n}\sqrt{E_C(2E-E_C)}\ $ & $\ S_H={2\pi R\over n}\sqrt{E_{BH}(2E-E_{BH})}\ $ \\
& 
\\
\hline
& 
\\
$\ E_C\equiv n(E+pV-TS)\ $ & $\ E_{BH}=n(E+pV -T_HS_H)\ $\\
& 
\\
\hline\hline
\end{tabular}

\medskip
{\it Table 2: Matching of the CFT-formulas with the FRW-equations} 
\end{center}
\medskip
\noindent
The presented relation between the FRW equations and the entropy formulas precisely holds at this transition point, when the holographic bound is saturated or
threatens to be violated. The miraculous merging of the CFT and FRW equations strongly indicates that both sets of these equations arise from a single underlying fundamental theory.

\medskip

The discovered relation between the entropy, Casimir
energy and temperature of the CFT and their cosmological counterparts
has a very natural explanation from a RS-type brane-world scenario \cite{rs} along the lines of \cite{gubser}. The
radiation dominated FRW equations can be  obtained by studying a brane
with fixed tension in the background of a
AdS-black hole. In this description the radius of the universe is
identified with the distance of the brane to the center of the black
hole. At the Big Bang the brane originates from the past singularity.
At some finite radius determined by the energy of the black hole, the
brane  crosses the horizon. It keeps moving away from the black hole,
until it reaches a maximum distance, and then it falls back into the
AdS-black hole. The special moment when the brane crosses the horizon
precisely corresponds to the moment when the cosmological entropy
bounds are saturated. This world-brane perspective on the cosmological
bounds for a radiation dominated universe will be described in detail
in \cite{ivo}.

 We have restricted our attention to matter described by a CFT in order
to make our discussion as concrete and coherent as possible.  Many of
the used concepts, however, such as the entropy bounds, the notion of a
non-extensive entropy, the matching of the FRW equations, and possibly
even the Cardy formula are quite independent of the equation of state
of the matter. One point at which the conformal invariance was used is
in the diagrammatic representation of the bounds. The diagram is only
circular when the energy $E$ goes like $R^{-1}$. But it is possible
that a similar non-circular diagram exists for other kinds of matter.
It would be interesting to study other examples in more detail.

 Finally, the cosmological constant has been put to zero, since only in
that case all of the formulas work so nicely. It is possible to modify
the formalism to incorporate a cosmological constant, but the analysis
becomes less transparent. In particular, one finds that the Hubble
entropy bound needs to be modified by replacing $H$ with the square
root of $H^2-\Lambda/n$. At this moment we have no complete
understanding of the case $\Lambda\neq 0$, and postpone its discussion
to future work. \bigskip

 \noindent {\sc{Acknowledgements}}

 \noindent I like to thank T. Banks, M. Berkooz, S. Gubser, G.
Horowitz, I. Klebanov, P. Kraus, E. Lieb, L. Randall, I. Savonije, G. Veneziano,
and H. Verlinde for helpful discussions. I also thank the theory
division at Cern for its hospitality, while this work was being
completed. \newpage

 \renewcommand{\theequation}{A.\arabic{equation}}
\setcounter{equation}{0}

 \end{document}